\documentclass[aps,prl,twocolumn,groupedaddress,showpacs]{revtex4}
\usepackage{color}
\usepackage{bm}
\usepackage{graphicx}%
\usepackage{amsmath}
\newcommand{\ket}[1]{\ensuremath{\left|{#1}\right\rangle}}
\newcommand{\vect}[1]{\ensuremath{\bm{#1}}}
\newcommand{\bra}[1]{\ensuremath{\left\langle{#1}\right |}}
\newcommand{\oper}[1]{\bm{\mathsf{#1}}}

\newcommand{\q}{\ensuremath{{\mathrm{q}}}}
\newcommand{\kv}{\ensuremath{{\mathrm{k}}}}

\newcommand{\bsy}[1]{\ensuremath{\boldsymbol{#1}}}
\newcommand{\brm}[1]{\ensuremath{\mathbf{#1}}}

\newcommand{\sinc}[1]{\ensuremath{\mathrm{sinc}\, #1}}
\newcommand{\vac}{\textsc{vac}}

\begin{document}

\title{Multimode Hong-Ou-Mandel interference}
\author{S. P. Walborn}
\email[]{swalborn@fisica.ufmg.br}
\affiliation{Universidade Federal de Minas Gerais, Caixa Postal 702, Belo
Horizonte, MG 30123-970, Brazil}
\author{A. N. de Oliveira}
\affiliation{Universidade Federal de Minas Gerais, Caixa Postal 702, Belo
Horizonte, MG 30123-970, Brazil}
\author{S. P\'adua}
\affiliation{Universidade Federal de Minas Gerais, Caixa Postal 702, Belo
Horizonte, MG 30123-970, Brazil}
\author{C. H. Monken}
\affiliation{Universidade Federal de Minas Gerais, Caixa Postal 702, Belo
Horizonte, MG 30123-970, Brazil}
\date{\today}
\begin{abstract} We consider multimode two-photon interference at a beam
splitter by photons created by spontaneous parametric down-conversion. The
resulting interference pattern is shown to depend upon the transverse spatial
symmetry of the pump beam. In an experiment, we employ the first-order
Hermite-Gaussian modes in order to show that, by manipulating the pump beam, one can control the resulting two-photon interference behavior. We expect these results to play an important role in the engineering of quantum states of light for use in quantum information processing and quantum imaging.
\end{abstract}
\maketitle
Entangled photons have proven to be a great tool in the
study of quantum phenomena and promise to play an important role in quantum
information processing \cite{chuang00} as well as quantum imaging
\cite{abouraddy01,strekalov02}. The most common source of entangled
photons is spontaneous parametric down-conversion (SPDC), in which the
interaction of a pump photon with a birefringent nonlinear crystal creates two
daughter photons. Under certain experimental
conditions, the down-converted photons may be entangled in momentum
\cite{rarity90b}, energy \cite{tapster94}, polarization \cite{kwiat95,kwiat99}
and/or angular momentum \cite{mair01}.
\par
Two-photon interference at a beam
splitter was first demonstrated by Hong, Ou and
Mandel (HOM) \cite{hom87}. It has since been utilized in quantum tests of
nonlocality \cite{torgerson95b} as well as many optical implementations of
quantum information protocol such as Bell-state measurements
\cite{mattle96,bouwmeester97} and may be used to construct quantum optical logic gates
\cite{ralph02,pittman02}. To date, however, most experiments 
utilizing HOM-type interference consider an ideal
monomode situation. In this paper, we consider multimode two-photon interference
of photon pairs created by SPDC. We show how the
transverse amplitude profile of the pump beam in SPDC determines whether the
down-converted fields interfere constructively or destructively.
We present our experiment and conclude by noting the relevance of these results to quantum optical information
processing.
\begin{figure}
\includegraphics[width=7cm]{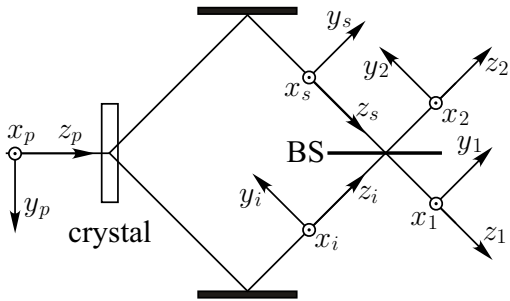}%
\caption{\label{fig:HOM}
  HOM interferometer.  SPDC-created photons are reflected onto a beam splitter (BS).}
\end{figure}

Consider the Hong-Ou-Mandel (HOM) interferometer shown in Fig. \ref{fig:HOM}, in which two photons are generated by SPDC are reflected onto opposite sides of a beam
splitter. We assume that paths $s$ and $i$ are equal.  Here we will work in the monochromatic and paraxial approximations.  This is justified by the use of narrow bandwidth interference filters (centered at $2\lambda_{p}$, where $\lambda_{p}$ is the pump field wavelength) and small collection apertures in the experimental setup. Following \cite{hong85,monken98a}, the two-photon quantum state generated by
noncolinear SPDC is then
\begin{equation}
\ket{\psi}_{SPDC} = C_{1}\ket{\vac} + C_{2}\ket{\psi}
\end{equation}
where
\begin{equation}
\ket{\psi}=\sum_{\sigma_{s},\sigma_{i}}C_{\sigma_{s},\sigma_{i}}\int\hspace{-2mm}\int\limits_{D}\hspace{-1mm} 
d\brm{q}_{s}
d\brm{q}_{i}\ \Phi(\brm{q}_{s},\brm{q}_{i})\ket{\brm{q}_{s},\sigma_{s}}_{s}
\ket{\brm{q}_{i},\sigma_{i}}_{i}.
\label{eq:14}
\end{equation}
The coefficients $C_1$ and $C_2$ are such that $|C_{2}| \ll \, |C_{1}|$. $C_2$
depends on the crystal length, the nonlinearity coefficient and the 
magnitude of
the pump field, among other factors.  The kets 
$\ket{\brm{q}_{j},\sigma_{j}}$ represent Fock states labeled by the 
transverse component $\brm{q}_{j}$ of the wave vector
$\brm{k}_{j}$ and by the polarization $\sigma_{j}$ of the mode $j = 
s$ or $i$.  The polarization state of the down-converted photon pair 
is defined by the coefficients $C_{\sigma_{s},\sigma_{i}}$. The 
function $\Phi(\brm{q}_{s},\brm{q}_{i})$, which can be
regarded as the normalized angular spectrum of the two-photon field
\cite{monken98a}, is given by
\begin{equation}
\Phi(\brm{q}_{s},\brm{q}_{i}) =\frac{1}{\pi}\sqrt{\frac{L}{K}}\
v(\brm{q}_{s}+\brm{q}_{i})\
\sinc\left(\frac{L|\brm{q}_{s}-\brm{q}_{i}|^{2}}{4K} \right),
\label{eq:15}
\end{equation}
where $v(\brm{q})$ is the normalized angular spectrum of the pump beam, $L$ is
the length of the nonlinear crystal in the $z$-direction, and $K$ is the
magnitude of the pump field wave vector. The integration domain $D$ is, in
principle, defined by the conditions $q_{s}^{2}\le k_{s}^{2}$ and $q_{i}^{2}\le
k_{i}^{2}$. However, in most experimental conditions, the domain in which
$\Phi(\brm{q}_{s},\brm{q}_{i})$ is appreciable is much smaller.  Eqs. (\ref{eq:14}) and (\ref{eq:15}) include the wave vectors inside the nonlinear birefringent crystal, which, upon propagation through the crystal, may suffer transverse and longitudinal walk-off effects, as well as refraction at the exit surface.  Walk-off effects can be corrected by  compensating crystals \cite{kwiat95}.  In the monochromatic situation we are considering, Snell's law gives equal exit angles for extraordinary and ordinary polarization.  Under these conditions, effects due to the refractive indices and birefringence can be neglected.  If the crystal is thin enough, the sinc
function in (\ref{eq:15}) can be considered to be equal to unity \cite{monken98a}. 
\par
The two-photon detection amplitude, which in the monochromatic case 
can be regarded as a photonic
wavefunction \cite{mandel95}, is
\begin{equation}
\bsy{\Psi}(\brm{r}_1,\brm{r}_2) =
\bra{\vac}\brm{E}_{2}^{(+)}(\brm{r}_2)\brm{E}_{1}^{(+)}(\brm{r}_1)\ket{\psi},
\label{eq:wf}
\end{equation}
where $\brm{E}_{l}^{(+)}(\brm{r}_{l})$ is the field operator for the mode $l$ and $\brm{r}_{l}$ is the detection position. In
the paraxial approximation, $\brm{E}_{l}^{(+)}(\brm{r})$ is
\begin{equation}
\brm{E}_{l}^{(+)}(\brm{r}) = e^{ikz} \sum\limits_{\sigma}\int d\brm{q}
\,\oper{a}_{l}(\brm{q},\sigma)\vect{\epsilon}_{\sigma}e^{i(\brm{q} \cdot
\vect{\rho}-\frac{\q^{2}}{2\kv}z)}
\label{eq:field}
\end{equation}
  The operator $\oper{a}_{l}(\brm{q},\sigma)$ annihilates a photon in 
mode $l$ with transverse
wave vector $\brm{q}$ and polarization $\sigma$.
\par
\begin{figure}
\includegraphics[width=5cm]{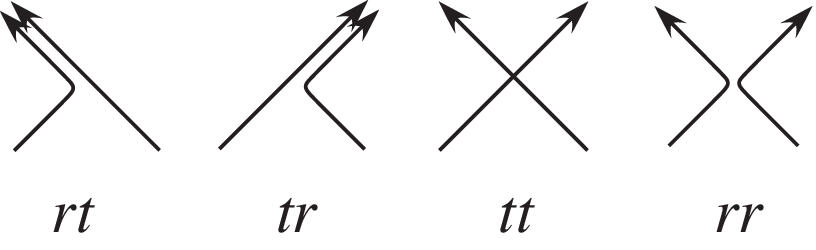}%
\caption{\label{fig:split}
  Possibilities of two-photon transmission and reflection.}
\end{figure}

In the HOM interferometer, the state (\ref{eq:14}) is incident on a beam
splitter. Using the reference frames illustrated in Fig. \ref{fig:HOM}, the annihilation operators 
in modes $1$
and $2$ after the beam splitter can be expressed in terms of the operators in
modes $s$ and $i$:
\begin{align}
\oper{a}_{1}(\brm{q},\sigma)& = t\oper{a}_{s}(q_{x},q_{y},\sigma) + i r
\oper{a}_{i}(q_{x},-q_{y},\sigma) \label{eq:aa} \\
\oper{a}_{2}(\q,\sigma)& = t\oper{a}_{i}(q_{x},q_{y},\sigma) + i r
\oper{a}_{s}(q_{x},-q_{y},\sigma),
\label{eq:ab}
\end{align}
where $t$ and $r$ are the transmission and reflection coefficients of the beam
splitter.  We have assumed that the beam splitter is symmetric.
A field reflected from the beam splitter undergoes a reflection in the horizontal ($y$) direction, while a transmitted field does not suffer any reflection, as illustrated in Fig. \ref{fig:HOM}. The negative sign that appears in the $q_{y}$ components is due to this
reflection.
The two-photon wave function is split into 
four components, according to the four possibilities of transmission 
and reflection of the two photons, as shown in Fig. \ref{fig:split}:
\begin{equation}
\label{eq:refl}
\bsy{\Psi} = \bsy{\Psi}_{tr}(\brm{r}_{1},\brm{r}^{\prime}_{1}) +
\bsy{\Psi}_{rt}(\brm{r}_{2},\brm{r}^{\prime}_{2}) +
\bsy{\Psi}_{tt}(\brm{r}_{1},\brm{r}_{2}) +
\bsy{\Psi}_{rr}(\brm{r}_{1},\brm{r}_{2}).
\end{equation}
For convenience, the four components of $\bsy{\Psi}$ are written in 
two different coordinate systems, $\brm{r}_{1}=(x_{1},y_{1},z_{1})$ and 
$\brm{r}_{2}=(x_{2},y_{2},z_{2})$, since we must work in the paraxial approximation 
around two different axes $z_{1}$ and $z_{2}$. To simplify things, we 
assume that $t=r$.  Combining Eqs. (\ref{eq:14}) through 
(\ref{eq:refl}), it is straightforward to show that, apart from a 
common multiplicative factor,
\begin{align}
\bsy{\Psi}_{tr}(\brm{r}_{1},\brm{r}^{\prime}_{1})=&
i \exp{\left\{\frac{iK}{2Z}\left[(x_{1}-x^{\prime}_{1})^{2}+(y_{1}+
y^{\prime}_{1})^{2}\right]\right\}}
\nonumber \times \label{eq:psitr}\\
&\left[\mathcal{W}\left(\frac{x_{1}+x^{\prime}_{1}}{2},\frac{-y_{1}+y^{\prime}_{1}}{2},Z\right)
  \bsy{\Pi}(\bsy{\sigma}_{1},\bsy{\sigma}^{\prime}_{1})+\right.\nonumber \\
&\left.\mathcal{W}\left(\frac{x_{1}+x^{\prime}_{1}}{2},\frac{y_{1}-y^{\prime}_{1}}{2},Z\right)
  \bsy{\Pi}(\bsy{\sigma}^{\prime}_{1},\bsy{\sigma}_{1})\right]
\end{align}
\begin{align}
\bsy{\Psi}_{rt}(\brm{r}_{2},\brm{r}^{\prime}_{2})=&
i \exp{\left\{\frac{iK}{2Z}\left[(x_{2}-x^{\prime}_{2})^{2}+
(y_{2}+y^{\prime}_{2})^{2}\right]\right\}}
\nonumber \times \label{eq:psirt}\\
& \left[ 
\mathcal{W}\left(\frac{x_{2}+x^{\prime}_{2}}{2},\frac{-y_{2}+y^{\prime}_{2}}{2},Z\right)\ 
\bsy{\Pi}(\bsy{\sigma}_{2},\bsy{\sigma}^{\prime}_{2})+\right.\nonumber \\
&\left. 
\mathcal{W}\left(\frac{x_{2}+x^{\prime}_{2}}{2},\frac{y_{2}-y^{\prime}_{2}}{2},Z\right)\ 
\bsy{\Pi}(\bsy{\sigma}^{\prime}_{2},\bsy{\sigma}_{2})\right]
\end{align}
\begin{align}
\bsy{\Psi}_{tt}(\brm{r}_{1},\brm{r}_{2})=&
\exp{\left\{\frac{iK}{2Z}\left[(x_{1}-x_{2})^{2}+(y_{1}-y_{2})^{2}\right]\right\}}
\nonumber \times\label{eq:psitt}\\
&\mathcal{W}\left(\frac{x_{1}+x_{2}}{2},\frac{y_{1}+y_{2}}{2},Z\right)\
\bsy{\Pi}(\bsy{\sigma}_{1},\bsy{\sigma}_{2})
\end{align}
\begin{align}
\bsy{\Psi}_{rr}(\brm{r}_{1},\brm{r}_{2})=&
- \exp{\left\{\frac{iK}{2Z}\left[(x_{1}-x_{2})^{2}+(y_{1}-y_{2})^{2}\right]\right\}}
\nonumber \times\label{eq:psirr}\\
&\mathcal{W}\left(\frac{x_{1}+x_{2}}{2},\frac{-y_{1}-y_{2}}{2},Z\right)\
\bsy{\Pi}(\bsy{\sigma}_{2},\bsy{\sigma}_{1})
\end{align}
where $K = 
k_{1}+k_{2}$ and,for simplicity, we consider $Z = z_{1}=z_{2}$. $\bsy{\Pi}(\bsy{\sigma}_{1},\bsy{\sigma}_{2})$ 
describes the polarization state of the photon pair.  $\mathcal{W}(x,y,Z)$ is the transverse field amplitude of the pump 
beam on the plane $z=Z$, which has been transferred to the two-photon wave function \cite{monken98a}. 
It is obvious that only $\bsy{\Psi}_{tt}$ 
and $\bsy{\Psi}_{rr}$ can give rise to coincidence detection in 
$D_{1}$ and $D_{2}$, while $\bsy{\Psi}_{tr}$ and $\bsy{\Psi}_{rt}$ 
correspond to two photons in arm $1$ and $2$, respectively.
 \par
 We now show how $\mathcal{W}$ and $\bsy{\Pi}$ affect the two-photon interference behavior.
Suppose that the photon pair is prepared in a symmetric polarization 
state: $\bsy{\Pi}(\bsy{\sigma}_{1},\bsy{\sigma}_{2}) =
\bsy{\Pi}(\bsy{\sigma}_{2},\bsy{\sigma}_{1})$.  If 
$\mathcal{W}(x,y,Z)$ is an even function with respect to the 
$y$-coordinate, that is, all the spatial components of $\bsy{\Psi}$ 
are symmetric wavefunctions, then Eqs. (\ref{eq:psitt}) and 
(\ref{eq:psirr}) cancel out and there can be no coincidence 
detections, as is well known \cite{hom87}.  However, if 
$\mathcal{W}(x,y,Z)$ is an odd function with respect to the 
$y$-coordinate, direct examination of  (\ref{eq:psitr}) through 
(\ref{eq:psirr}) shows that (\ref{eq:psitr}) and (\ref{eq:psirt}) are 
zero, while (\ref{eq:psitt}) and (\ref{eq:psirr}) add constructively, 
resulting in an increase in coincidence counts.
Now suppose that the photon pair is prepared in an antisymmetric 
polarization state: $\bsy{\Pi}(\bsy{\sigma}_{1},\bsy{\sigma}_{2}) = -
\bsy{\Pi}(\bsy{\sigma}_{2},\bsy{\sigma}_{1})$.   Then  for a 
$\mathcal{W}(x,y,Z)$ that is an even function of $y$, clearly 
Eqs. (\ref{eq:psitr}) and (\ref{eq:psirt}) are zero, while 
(\ref{eq:psitt}) and (\ref{eq:psirr}) add constructively, giving
 an increase in the coincidence counts.  For $\mathcal{W}(x,y,Z)$ 
that is an odd function of $y$, (\ref{eq:psitt}) and (\ref{eq:psirr}) 
cancel, eliminating coincidence detections.
\par
The behavior of the HOM interferometer for any combination of 
symmetric and antisymmetric spatial and polarization components of 
$\bsy{\Psi}$ can be inferred from the bosonic character of photons, 
that is $\bsy{\Psi}$ must be symmetric.
\par
To our knowledge, all HOM-type experiments performed up until now have used a 
pump beam that is described by an even function  of $y$.  In order to 
demonstrate experimentally the possibilities of controlling the HOM 
interferometer with space and polarization variables, we performed a 
series of experiments in which coincidence counts were registered, 
combining symmetric and antisymmetric components of $\bsy{\Psi}$.
\par
A a set of beams with well defined cartesian parity are the 
Hermite-Gaussian (HG) beams, given by
\cite{beijersbergen93}
\begin{align}
\mathcal{W}_{mn}(x,y,z) = & C_{mn}
H_{m}(x\sqrt{2}/w)H_{n}(y\sqrt{2}/w) e^{(x^2+y^2)/w^2}  \nonumber \\
& \times e^{-ik(x^2+y^2)/2R} e^{-i(m+n+1)\theta} \nonumber 
\end{align}
where $C_{mn}$ is a constant. The $H_{n}(y)$ are the Hermite polynomials, which
are even or odd functions in the $y$-coordinate when the index $n$ is even
or odd, respectively. $w$ is the beam waist,
$R(z) = (z^2+z_{R}^2)/z$ and
$\theta(z) =\arctan(z/z_{R})$,
where $z_{R}$ is the Rayleigh range.
\par
Pumping the nonlinear crystal with different HG pump 
beams, we can
control the behavior of the down-converted photons at the beam splitter.
\begin{figure}
\includegraphics[width=7cm]{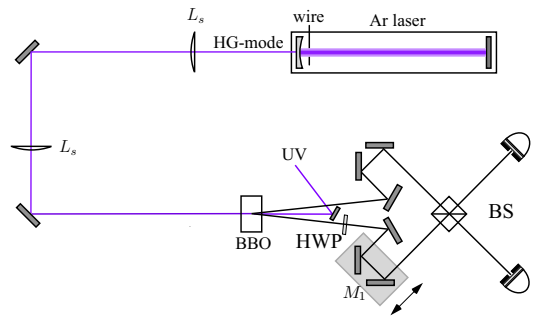}%
\caption{\label{fig:exp}
  Experimental setup.  The Argon laser is used to pump a  BBO
($\beta$-BaB$_2$O$_4$) crystal cut for degenerate type II phase matching,
generating noncollinear entangled photons by spontaneous parametric
down-conversion (SPDC).  The wire is used to generate HG modes (see text).  The $L_{s}$ are spherical lenses ($f=500\,$mm) used to
focus the pump beam in the plane of the detectors to increase the coincidence
detection efficiency \cite{monken98b}. The down-converted photons are reflected
through a system of mirrors and incident on a $50-50$ beam splitter BS ($t =r
\approx \sqrt{1/2}$).  A computer-controlled stepper motor
is used to to control the path length difference by scanning mirror assembly $M_{1}$.   The detectors $D_{1}$ and $D_{2}$
are free space EG{\&}G SPCM 200 photodetectors, equipped with interference 
filters ($1\,$nm
FWHM centered at $702\,$nm) and $2\,$mm circular apertures. Coincidence and
single counts were registered using a personal computer.}
\end{figure}
The experimental setup is a typical HOM interferometer \cite{hom87}, 
shown in Fig.
\ref{fig:exp}.  To generate HG modes we placed a $25\,\mu$m 
diameter wire inside
the laser cavity, forcing the laser to operate in one of the HG modes with a nodal line at the position of the wire \cite{beijersbergen93}.  The wire is mounted on a rotational stage.  We were able to
generate first-order modes in any direction ($x,y,\pm 45$, etc.) with laser
power $\sim 30$\,mW.  Symmetric and antisymmetric polarization states 
were used.
The symmetric state chosen was $\ket{\Pi^{S}} = 
\ket{H}_{1}\ket{H}_{2}$ and the antisymmetric state was 
$\ket{\Pi^{A}} = 
\frac{1}{\sqrt{2}}(\ket{H}_{1}\ket{V}_{2}-\ket{V}_{1}\ket{H}_{2})$, 
where $H$ and $V$ stand for horizontal and vertical linear 
polarization, respectively.
The  state $\ket{\Pi^{S}}$ was obtained from 
type II SPDC, by collecting 
one photon from the  ordinary 
($H$-polarized) light cone and the other photon from the extraordinary 
($V$-polarized) light cone followed by a half-wave plate, which 
rotates its polarization to $H$. Realigning the crystal, the antisymmetric state 
$\ket{\Pi^{A}}$ was obtained from the crossing of the ordinary and 
extraordinary light cones, followed by compensators, as described in 
Ref. \cite{kwiat95}.
\par
\begin{figure}
\includegraphics[width=6cm]{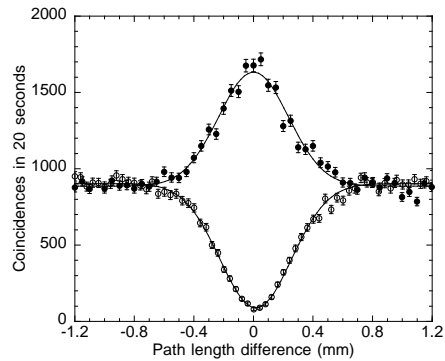}%
\caption{\label{fig:sym}  Coincidence counts when the polarization 
state is symmetric and the pump beam is a first-order 
Hermite-Gaussian beam. Open circles ($\circ$) correspond to 
$\mathcal{W}_{10}$ and  solid circles ($\bullet$) correspond to 
$\mathcal{W}_{01}$.}
\end{figure}
\begin{figure}
\includegraphics[width=6cm]{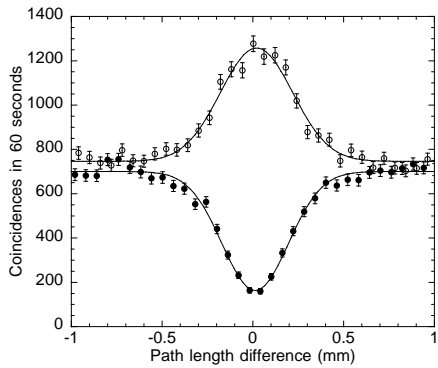}%
\caption{\label{fig:asym}  Coincidence counts when the polarization 
state is antisymmetric and the pump beam is a first-order 
Hermite-Gaussian beam. Open circles ($\circ$) correspond to 
$\mathcal{W}_{10}$ and  solid circles ($\bullet$) correspond to 
$\mathcal{W}_{01}$.}
\end{figure}
\begin{figure}
\includegraphics[width=6cm]{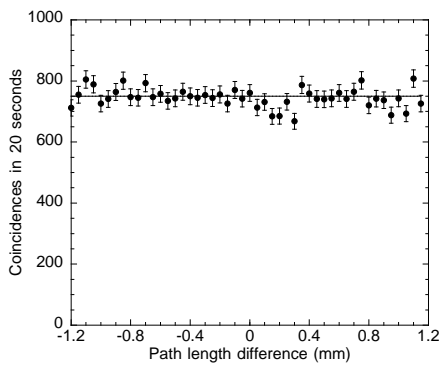}%
\caption{\label{fig:comb}  Coincidence counts when the polarization 
state is symmetric and the pump beam is an equal superposition of 
Hermite-Gaussian modes
$\mathcal{W}_{01}$ and $\mathcal{W}_{10}$.}
\end{figure}

Experimental results are sumarized in Figs. \ref{fig:sym}, 
\ref{fig:asym} and \ref{fig:comb}.  The error bars represent statistical errors due to photon counting \cite{mandel95}.
\par
Fig. \ref{fig:sym} shows the results for the symmetric polarization 
state $\ket{\Pi^{S}}$ when the crystal is pumped by first-order 
HG
beams $\mathcal{W}_{10}$ and $\mathcal{W}_{01}$. $\mathcal{W}_{10}$, 
as an even function
in $y$, results in $\bsy{\Psi}_{rr} = -\bsy{\Psi}_{tt}$, leading to 
an interference minimum. $\mathcal{W}_{01}$, as an odd function in 
$y$, results in $\bsy{\Psi}_{rt} = \bsy{\Psi}_{tr} = 0$ and
$\bsy{\Psi}_{rr} = \bsy{\Psi}_{tt}$, leading to an interference maximum.  The curves have visibilities of $0.90\pm0.01$ ($\mathcal{W}_{10}$) and $0.85\pm0.01$ ($\mathcal{W}_{01}$).
Fig. \ref{fig:asym} shows the results for the antisymmetric 
polarization state $\ket{\Pi^{A}}$ under the same pump beam 
conditions. Now, the behavior of the interference is the opposite, 
that is, pumping with $\mathcal{W}_{10}$ produces an interference 
maximum, whereas pumping with $\mathcal{W}_{01}$ produces a minimum.  Visibilities of $0.70\pm0.01$ ($\mathcal{W}_{10}$) and $0.77\pm0.01$ ($\mathcal{W}_{01}$) were achieved.  Differences in the visibilites were due to the alignment of the interferometer as well as the wire in the laser cavity.    
We can create an equally weighted superposition of the
modes $\mathcal{W}_{01}$ and $\mathcal{W}_{10}$ by placing the wire in the
cavity at a $45^\circ$ angle  \cite{beijersbergen93}. Such a superposition is neither symmetric nor antisymmetric.  As to be expected, the coincidence count rate is
constant, as shown in Fig. \ref{fig:comb} for the symmetric polarization state.
\par
We have theoretically and experimentally investigated mulitmode HOM
interference using photons generated by SPDC. The resulting
interference pattern is seen to depend upon both the transverse 
amplitude profile of
the pump laser and the polarization state of the photon pair. 
We used 
first-order Hermite-Gaussian beams to demonstrate that
by manipulating the pump beam, one can control the two-photon interference.
To our knowledge,
this is the first time that two-photon HOM interference has been studied using
a spatially antisymmetric wave function.  
\par
A possible application of these results is Bell-state analysis without the need for detectors sensitive to photon number \footnote{Such an experiment is currently underway in
our laboratory.}.  We
expect these interference effects to be important in the construction of 
quantum-optical logic gates 
\cite{klm01, ralph02,pittman02}, as well as the codification of information in the transverse spatial properties
of the photon \cite{leach02}.
In addition, there is the possibility of using these results to create non-classical states of light with spatial properties that could be useful for 
quantum imaging.
\begin{acknowledgments}
The authors thank the Brazilian funding agencies CNPq and CAPES.
\end{acknowledgments}

\end{document}